\documentclass{emulateapj}
\usepackage{times}
\usepackage{graphicx}% Include figure files

\slugcomment{To Appear in ApJ Letters}

\shorttitle{2FGL J1311.7$-$4329}
\shortauthors{Romani}

\begin{document}
%\linenumbers

\title{2FGL J1311.7$-$3429 Joins the Black Widow Club}

%\author{R. W. Romani\affilmark{1}, M. Kerr\altaffilmark{1}, H. A. Craig\altaffilmark{1}, 
\author{Roger W. Romani\altaffilmark{1}}% and  M. Kerr\altaffilmark{1}}
\affil{Department of Physics, Stanford University, Stanford, CA 94305} 
\altaffiltext{1}{Visiting Astronomer, Kitt Peak National Observatory and Cerro Tololo InterAmerican Observatory, National Optical Astronomy Observatory, which is operated by the Association of Universities for Research in Astronomy (AURA) under cooperative agreement with the National Science Foundation. The WIYN Observatory is a joint facility of the University of Wisconsin-Madison, Indiana University, Yale University, and the National Optical Astronomy Observatory.
}

%\altaffiltext{1}{Einstein Fellow}

\begin{abstract}

	We have found an optical/X-ray counterpart candidate for
the bright, but presently unidentified, {\it Fermi} source 
2FGL J1311.7$-$3429.  This counterpart undergoes large amplitude quasi-sinusoidal optical 
modulation with a 1.56h (5626s) period. The modulated flux is
blue at peak, with $T_{\rm eff} \approx$14,000\,K, and redder at minimum. 
Superimposed on this
variation are dramatic optical flares. Archival X-ray data suggest modest binary modulation,
but no eclipse. With the $\gamma$-ray properties, this
appears to be another black-widow-type millisecond pulsar. If confirmation
pulses can be found in the GeV data, this binary will have the shortest orbital period
of any known spin-powered pulsar. The flares may be magnetic events on
the rapidly rotating companion or shocks in the companion-stripping wind.
While this may be a radio-quiet millisecond pulsar, we show that
such objects are a small subset of the $\gamma$-ray pulsar population.
\end{abstract}


\section{Introduction}

In the most recent {\it Fermi} Large Area Telescope (LAT)
catalog of 1873 $0.1-100$\,GeV sources, over 1170 have statistically
reliable counterpart identifications at lower energy \citep{2FGL}. The bulk of
these identifications are blazars and spin-powered pulsars, with 
young pulsars (both radio-selected and Geminga-like $\gamma$-ray selected) dominating 
at low Galactic latitude.  An additional population of millisecond pulsars
extends to high latitude among the blazar-associated sources. 
Identification progress is especially impressive for the bright, well localized sources that
have been studied since early in the {\it Fermi} mission.
Here we update \citet{bsl}, selecting `bright' sources from the 2FGL catalog 
as those with $>20\sigma$ detection significance and time-averaged energy flux 
$F_E > 3 \times 10^{-11}{\rm erg\,cm^2\,s^{-1}}$ -- there are 249 such sources. 
All but six presently have lower energy identifications: blazars, pulsars 
and a few binaries.  The handful of sources {\it not} yet associated with one
of these source classes provide the best prospect for new types of 
$\gamma$-ray emitter. We have initiated a campaign to characterize these unidentified 
sources. The first result from this effort is the discovery of dramatic optical 
and X-ray variability for the high latitude ($|b|=62^\circ$) source J2339$-$0533, 
implying that it is a short period `black-widow'-type binary millisecond 
pulsar (MSP) \citep{rs11,ket12}.

	Here we report progress on the next high latitude unidentified
source in this set, 2FGL J1311.7$-$3429 (hereafter J1311) at $|b|=28^\circ$. With a $43\sigma$ detection
significance (the highest among the 2FGL unidentified sources) and an energy flux
of $F_{0.1-100GeV} = 6.2 \times 10^{-11}{\rm erg\,cm^{-2}\,s^{-1}}$ (the second brightest
UnIDed 2FGL), this is a top candidate for follow-up. The `Variability index' value
19.5 and `Curvature Significance'=6.3 mark this as a steady source with a substantial
spectral cut-off: a prime pulsar candidate. It has been searched for 
$< 30$\,Hz $\gamma$-ray pulsations \citep[eg.][]{blind} and for radio pulses
down to $\sim$\,ms periods \citep{ray12}, with no detection. Thus it is unlikely
to be an isolated young pulsar or a radio-loud MSP. We describe here an optical campaign to
identify and characterize a counterpart.

%	We have obtained optical photometry revealing large amplitude periodic modulation.
%Together with weaker evidence for X-ray modulation, the data suggest a 
%`black-widow' MSP binary. However, we find that this system has an exceptionally short
%1.5h orbital period and shows dramatic flares implying strong eruptive events in the system.
%As for J2339$-$0533, the strong evaporative wind and the short period binary may 
%prevent radio-detection of the MSP. We
%discuss initial models to characterize this binary and review briefly the prospects
%for {\it Fermi} detection and the implications for other bright unidentified $\gamma$-ray
%sources.

\section{Observations}

	Initial exposures of the J1311 error region were taken with the 
MiniMo camera at the 3.6\,m WIYN telescope on Feb. 17-18, 2012. Exposures
were 180\,s using the Gunn filter set ($2\times g,\,r,\,i$ on 2/17; 
$5\times g,\,r$ on  2/18). Observations were perforce at airmass $2.5-3$,
seeing was variable and conditions were non-photometric.

A more extensive image sequence was taken with the SOAR Optical Imager (SOI)
at the 4.2\,m SOAR telescope on March 21-23, 2012. Here 120\,s exposures
using SDSS filters were taken, covering one color each night 
($61\times g^\prime$ on 3/21, $74\times r^\prime$ on 3/22 and $57\times 
i^\prime$ on 3/23). The camera was binned $2\times2$, allowing fast 11\,s
read-out and high observation efficiency. Again image quality was highly
variable (partly due to loss of wavefront sensor control of the primary).

	We searched image archives for exposures covering J1311, finding
a sequence (program 086.D-0388) of 170\,s VLT/VIMOS exposures taken on March 4 
($2\times [3B,\,3V,\,3 I]$) and March 11, 2011 
($3\times [3B,\,3V,\,3I]$). VIMOS is an array camera 
with 2$^\prime$ gaps between the four CCD chips, so the five pointings were 
offset to cover the full J1311 error region.  

	In examining the images for variable objects, we paid particular
attention to sources coincident with detections in archival X-ray data (see below).
One such star was blue and highly variable on short timescales. The
NOMAD position of this likely counterpart is 13 11 45.741, -34 30 29.96 (2000.0).
While we ensured that this star was covered in all MiniMo and SOI exposures,
we found that the counterpart was in the array gap for the first VIMOS pointing on
March 3, 2011 and at the extreme edge of a chip, suffering major vignetting,
in the second pointing on March 11, 2011.

	To quantify the variations, we performed simple aperture photometry of
the counterpart and surrounding field stars in all frames. Since the bulk of the imaging
was in the SDSS filter set, we attempted to reference all photometry to this scale.
The MiniMo data were calibrated using $g\, r\, i$ images in SDSS fields
at low airmass; we transferred this calibration to stars in the J1311 field.
Unfortunately, the non-photometric WIYN conditions and pass-band differences between
the SDSS and Gunn sets make the substantial atmospheric extinction
corrections for J1311 uncertain. This is not a severe problem for the field stars,
which were mostly late type with $g^\prime-r^\prime=0.7-1.5$. However the target's color is
as blue as $g^\prime-r^\prime=-0.5$ and varies appreciably with magnitude. Thus
while the relative photometry for each color sequence has very small statistical 
errors, there may be a systematic offset between the WIYN and SOAR points as large 
as 0.2\,mag, especially in $g^\prime$. For the VIMOS images, we used NOMAD field
stars to establish zero points in B and V. NOMAD-detected stars were generally 
saturated in the VIMOS I images, so we established the I scale using stars from 
the DENIS catalog.

	We show the photometry for each night in Figure 1, converting to 
fluxes and plotting against estimated binary phase.
The SOI $g^\prime,\, r^\prime$ data show dramatic quasi-sinusoidal variability by 
a factor $\sim 30\times$ with a period of $\sim 1.5$\,h. In $i^\prime$ an initial 
flux minimum was followed by a large $\sim 5\times$ flare, with rapid fluctuations 
during return to quiescence. The  $g^\prime,\, r^\prime$ fluxes also show variability,
especially near maximum. The WIYN data are sparse and have larger statistical errors,
but do follow the periodic modulation. Most WIYN fluxes appear 10-20\% lower
than SOAR at the corresponding phase, suggesting an underestimate of $\sim 0.1$\,mag
for the extinction correction.  However, given the apparent non-periodic
variability this cannot be firmly concluded. The VLT fluxes from March 11 pass 
through two minima; those from March 4 show one minimum, but appear to have
$\sim 8\mu$Jy additional flux.

\begin{figure}[t!!]
\vskip 9.3truecm
\includegraphics{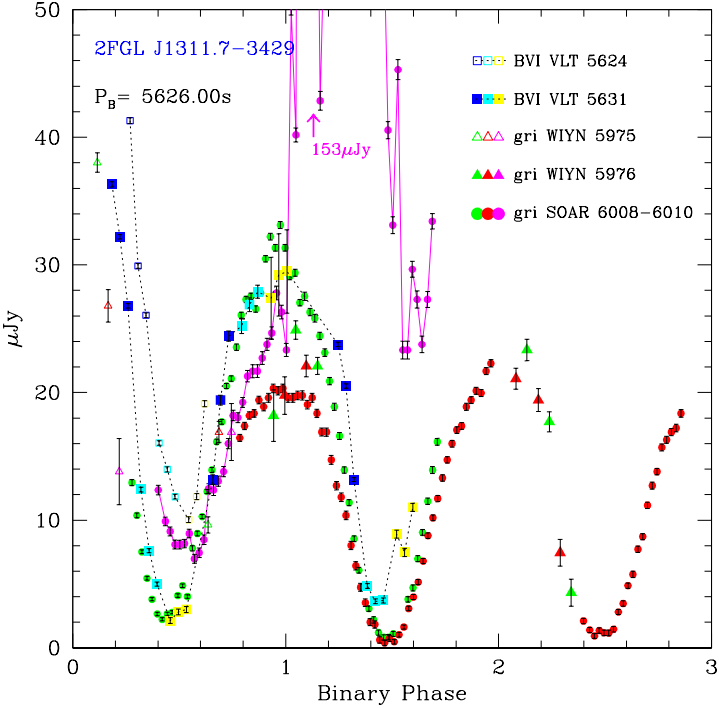}
\begin{center}
\caption{\label{LCfluxs} 
SOAR/WIYN/VLT photometry of J1311.7$-$3429. The abscissa is the phase relative to
phase 0 (superior conjunction, secondary flux maximum) just before the first
observation of each plotted night, using the best-fit $P_b=$5626.00\,s period. 
Symbols and colors indicate the various telescopes and observing bands. 
Photometry from the SOAR $i^\prime$ flare is connected with a solid line;
points for each VLT night are connected with a dash line.
}
\end{center}
\vskip -0.7truecm
\end{figure}

\subsection{Orbital Period Estimate}

	The intermittent flaring activity of J1311 compromises simple
periodogram or Fourier methods for refining the period. However, 
the combination of SOAR, WIYN and VLT exposures give photometry information on
1-3, 7, $\sim$35, and $\sim$350 day time scales, which suffices to resolve
possible aliases. After referencing all observation mid-point times to the 
solar system barycenter, we can constrain the phase of the 2012 epoch minima 
to $\delta \phi \approx 0.02$, giving a best fit period of $P_b=5626.00\pm0.02$s. 
The epoch of (quiescent) maximum light associated with our SOAR $r^\prime$ measurements
(presumably pulsar superior conjunction, see below) is %JD 2456009.6776 \pm 0.001
barycentric MJD\,$56009.1795 \pm 0.0013$.

The approximately sinusoidal nature of the light curve and very large modulation 
amplitude can be best interpreted in terms of strong companion heating. The
inferred heating luminosity is appreciably above any X-ray flux (see below); 
we infer that accretion power is not dominant and that the heating source is invisible
in the optical and faint in the X-ray. We conclude that this is a 
`black-widow' type pulsar
binary and is the likely counterpart of the $\gamma$-ray source. In this it is
remarkably similar to the black widow candidate counterpart for J2339$-$0533.
But the $\sim 3\times$ shorter orbital period is unprecedented; this is an
extreme black widow system.

\begin{figure*}[t!!]
\vskip 7.2truecm
\includegraphics{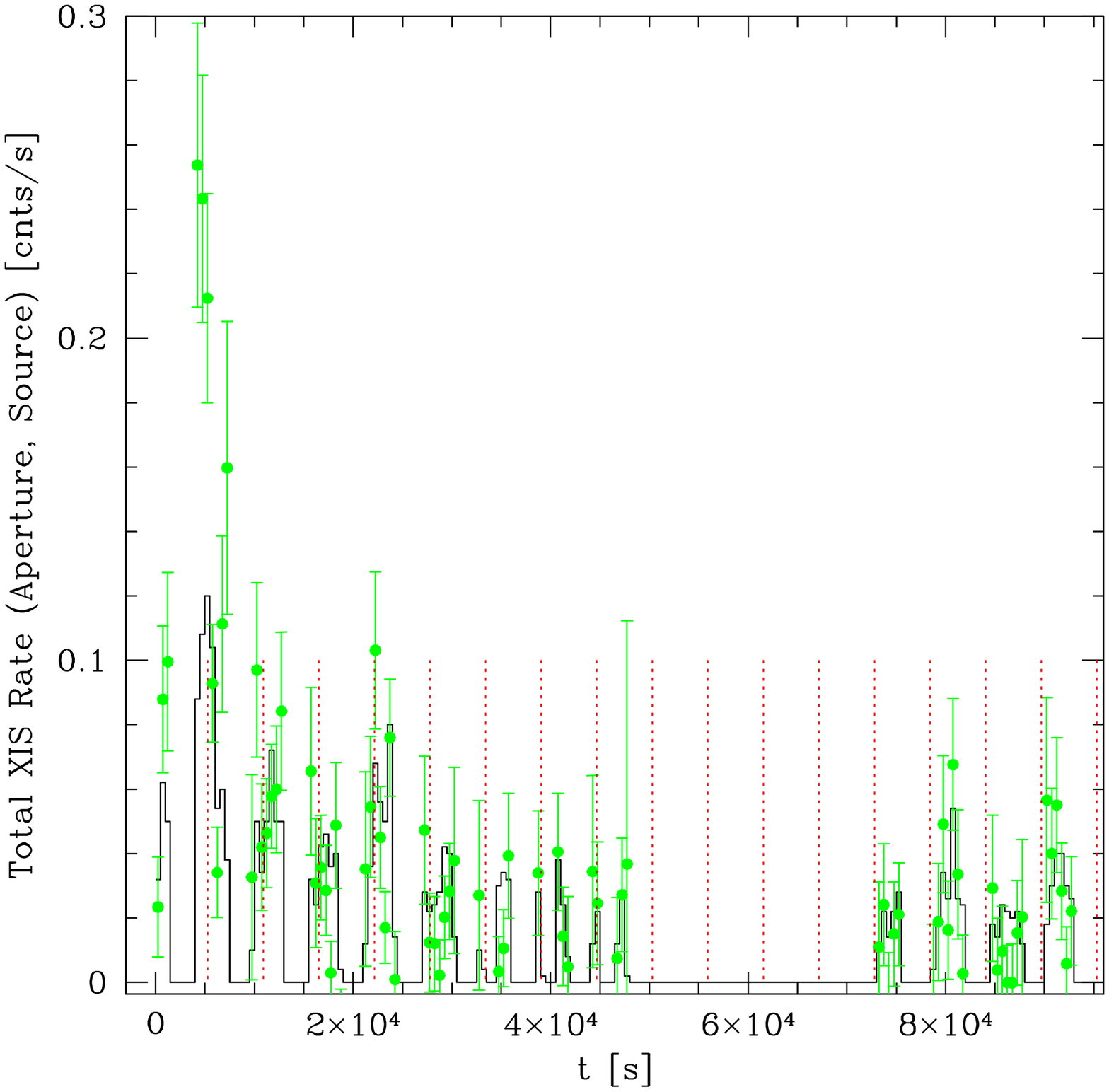}
\includegraphics{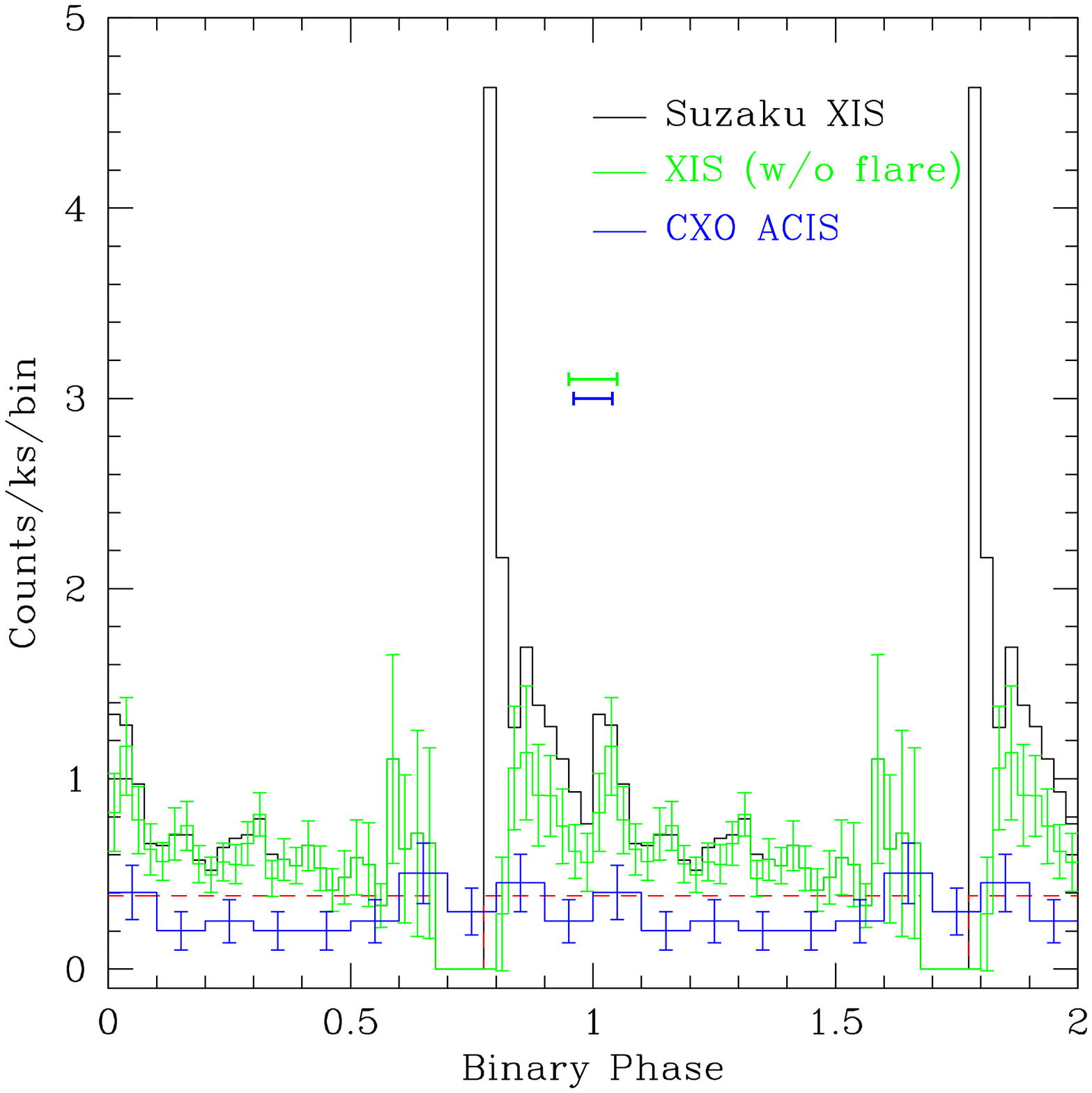}
\begin{center}
\caption{\label{SuzLC} Left: Raw aperture count rate (histogram)
and background- and aperture- corrected counts (points) during the Suzaku exposure.
Times of optical flux maximum are marked.
Right: Two periods of the X-ray orbital light curve of J1311.7$-$3429. Statistical 
errors are shown for the exposure-corrected bin count rate. For Suzaku, we show 
the folded data both with and without the bright flare. Horizontal error
bars show the extrapolated phase uncertainty.
}
\end{center}
\vskip -0.1truecm
\end{figure*}

\subsection{Archival X-ray Light Curve}

	2FGL J1311.7$-$3429 has been observed by both the {\it CXO} and {\it Suzaku}
satellites. One exposure from each facility is available in the NASA archives,
additional exposure has been taken but is not publicly available. The counterpart
was detected in both data sets and well localized by {\it CXO}. We analyzed both
observations for X-ray spectrum and variability.

	The {\it Suzaku} observation (ObsID 804018010; Kataoka, PI) started
on August 04 2009 (MJD 55047.2290) and had 33.4\,ks of on-source time over 95\,ks
($\sim 17$ binary orbits) with exposure in XIS0,1,3.  A comparably bright source lies
$\sim 1.6^\prime$ from the target, so we extracted source counts from a $1^\prime$ radius aperture.
Figure 2 (left) shows the combined XIS count rate, both direct aperture counts (histogram) and the 
background-subtracted aperture-corrected count rate (points).  The vertical 
dotted lines mark the phases of maximum optical light;
we see that orbital modulation is difficult to measure since $P_b$ is so close to
that of satellites in low Earth orbit. A clear X-ray flare is seen 5000\,s
after the observation start.

	An ACIS-I exposure of 19.8\,ks live-time (ObsID 11790; Cheung, PI) was 
taken starting MJD 5527.6652. Unfortunately, as for VLT observation 1, the
counterpart was directly in the gap between ACIS-I chips.  However, the 
{\it CXO} dither did provide some ($\sim$25\% effective live-time) exposure and a source
coincident with the optical counterpart was well detected with 60 counts. There 
is no strong evidence for 
secular X-ray variability on 3000\,s timescales. The exposure only covers $\sim 3.5$ 
orbits. Figure 2 (right)  shows the folded X-ray light curves from the {\it Suzaku} and
{\it CXO} data sets. After the early flare in the {\it Suzaku} data is excluded
there is only weak evidence for orbital modulation; note there is an exposure gap
for phases $\phi_B=0.7-0.8$. Similarly the {\it CXO} data show no strong peak or eclipse.
In both light curves there seems to be a slight excess at phases 0.7-1.1.
This may indicate additional weaker flaring activity near superior conjunction.

	Only crude spectral constraints can be obtained with the limited counts,
especially after selecting for {\it Suzaku} photons outside of the flare period.
If we fit an absorbed power law to the X-ray data we find that $N_H$ 
is essentially unconstrained ($0.5 \pm 6 \times 10^{21} {\rm cm^{-2}}$) but consistent with the limited
extinction in this high $|b|$ direction. We therefore fix at the 
$N_H = 5 \times 10^{20} {\rm cm^{-2}}$ inferred from the dust maps (HEASARC NH tool),
and fit for the power law index. Using CIAO the {\it CXO} data fit to
$\Gamma=1.3\pm0.5$. XSPEC fits to the {\it Suzaku} XIS0,3 give
$\Gamma=1.6\pm0.8$. Fixing at $\Gamma=1.5$, we obtain a quiescent unabsorbed 
flux for the source of 
$f_{0.3-10\,keV} = 2.4 \pm 0.6 \times 10^{-13}  {\rm erg\, cm^{-2}\, s^{-1}}$.
A deeper exposure, allowing a better light curve and a orbit-dependent
spectral study is clearly of interest.

\begin{figure}[t!!]
\vskip 9.15truecm
\includegraphics{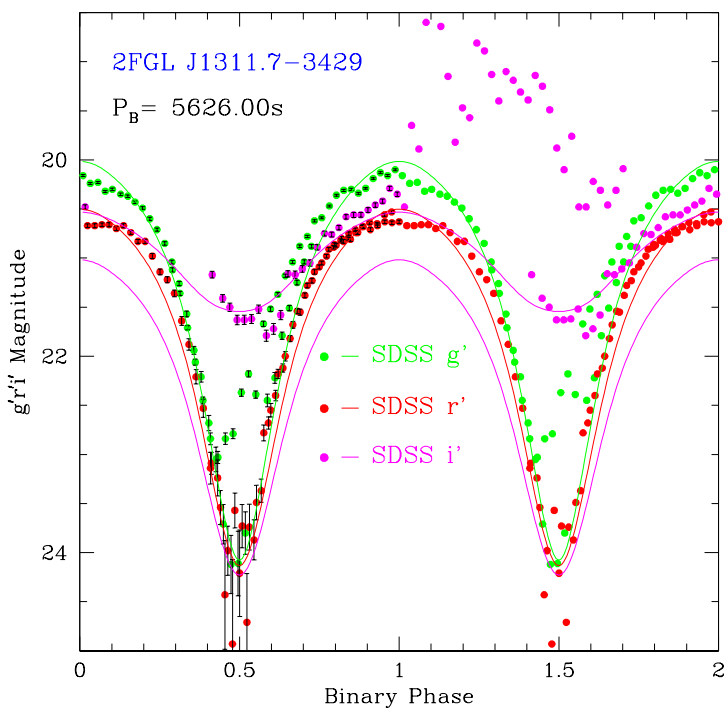}
\begin{center}
\caption{\label{LCmag} 
Folded SOAR $g^\prime r^\prime i^\prime$ light curves of J1311.7-3429 (two periods). 
Statistical errors are shown during the first period only and the large 
$i^\prime$ flare is plotted during the second period only. The curves represent
a simple heated secondary model from ELC; a second copy of the $i^\prime$
curve with a constant flux offset is also shown for comparison.
}
\end{center}
\vskip -0.7truecm
\end{figure}

\section{System Modeling}

	Figure 3 shows two periods of the SOI light curves folded on the
ephemeris of \S2.1.
The large quasi-sinusoidal modulation shows that the light curve is
dominated by heating of the secondary. The typical $g-r \approx -0.4$ near $\phi=0$
suggests $T_{\rm eff} \approx 14,000$\,K. By phase $\phi =0.35$ the color has reddened
to $g-r > -0.1$ ($T_{\rm eff} \approx 8000$\,K). At flux minimum ($\phi \sim 0.5$) large
(up to $\Delta m= 3$) fluctuations dominate. Since relatively small fluctuations
can overwhelm the quiescent flux, these prevent any meaningful measurement of color
at flux minimum. 

	We have attempted to model the `quiescent' light curves with the ELC
code \citep{oh00}. A formal fit is not indicated as there are many positive
`flare' excursions from any reasonable light curve. However comparison with 
a range of models shows that a few of the parameters are significantly constrained.
For example, matching the characteristic $g^\prime-r^\prime$ at maximum
requires heating by a primary luminosity (modeled here as an isotropic `X-ray' 
flux) of ${\rm Log}[L_x({\rm erg/s})] = 36.0 \pm 0.3$. Also, to obtain
adequate modulation while avoiding an X-ray eclipse, requires $60^\circ<i<80^\circ$.
At this point other parameters are poorly constrained, although large mass
ratios and a near-Roche-lobe filling secondary are preferred. However,
the basic ELC model seems inadequate for a good fit; a particular challenge 
is to reproduce the relatively flat maxima and broad `shoulders' in 
the $g^\prime$ and $r^\prime$ curves at $\phi \sim 0.25$, 0.75. This may
be due to limitations of the atmosphere tables. One can substantially improve
the fit by modifying the limb-darkening law, but for blackbody surface spectra
broad light curves (with adequate heating) require unphysical negative limb
darkening coefficients. Another, more attractive, possibility is that the
distribution of surface heating differs from the isotropic form assumed here;
the pulsar may primarily heat a portion of the companion star. An illustrative
set of curves is shown in Figure 3, assuming a mass ratio $q=50$, orbital
semi-major axis $a=0.9R_\odot$, inclination $i=65^\circ$, heating flux
$10^{36}{\rm erg\,s^{-1}}$ and a $T_{\rm eff}=4000$\,K companion reaching
0.99 of its Roche lobe radius. The observed $i^\prime$ flux evidently has a
substantial added flux even before the bright flare, at the epoch observed.

\section {Discussion}

	At this short orbital period, the mean density in the secondary
Roche lobe is $\langle \rho_2 \rangle = 46 {\rm g\,cm^{-3}}$, so 
the secondary could be a main sequence star of $< 0.15M_\odot$. To reach 
this period, however, the system very likely passed through a common envelope
phase, and would have a helium-rich core. The strong heating and
near-Roche lobe filling suggest a sub-stellar secondary, but additional
observations are needed to determine the state of the secondary core and envelope.

	A very rough estimate of the source distance can be made from the 
$g^\prime -r^\prime =-0.4$ color and $g^\prime = 20.2$  flux at maximum, if
we assume that the star nearly fills its Roche lobe at 
$\approx 0.46 a (M_2/M_{Tot})^{1/3} \approx 0.1 R_\odot$.
Such color indicates $T_{\rm eff} \approx$14,000\,K (type $\sim$\,B6), 
corresponding to main sequence $M_g =-0.43$
and $R\approx 3.0\,R_\odot$. Correcting for the effective area and the small
($A_g = 0.25$) Galactic extinction in this direction, we get 
$d \approx 3.9 (q/50)^{-1/3}$kpc.
This is likely an overestimate as the secondary should be H-poor,
and the temperature of the viewed hemisphere is far from uniform, even at maximum.

	If we adopt the view that the X- and $\gamma$-ray emissions come from an
energetic pulsar, we can make some additional estimates. \citet{beck09} finds that
pulsars have $L_X \approx 10^{-3}{\dot E}$, so from the observed X-ray flux we would
infer a spin-down luminosity ${\dot E} \approx (10^3) 4\pi d^2 2.4 \times 10^{-13} {\rm erg\,
cm^{-2}s^{-1}}\approx 2.9 \times 10^{34}d_{kpc}^2 {\rm erg\,s^{-1}}$. Gamma-ray pulsars are also
observed to follow a heuristic luminosity law 
$ L_{\gamma,heu} \approx ({\dot E}\, \times \, 10^{33}{\rm erg/s})^{1/2} $
\citep{psrcat}. This is related to the observed flux as
$ L_\gamma = 4\pi f_\Omega F_\gamma d^2, $
where the beaming-dependent correction factor is $f_\Omega \sim 0.7-1.3$ \citep{wet09}.
Thus the observed $F_\gamma = 6.2 \times 10^{-11} {\rm erg\,cm^{-1}\,s^{-1}}$ gives 
a distance-dependent estimate for the spindown power of 
${\dot E} \approx 5.5 \times 10^{34} f_\Omega^2 d^4 {\rm erg  \,s^{-1}}$. 
This is consistent with the X-ray luminosity estimate when $d\approx 0.75 f_\Omega^{-1}$\,kpc.
However, if we match ${\dot E}$ to the ELC-estimated heating flux a distance
of $\sim 2$\,kpc is preferred.

	The nature of the intermittent `flaring' emission is at present unclear. Since
the bands were observed sequentially, we do not even know the characteristic color.
All we can infer are large increases above the quiescent flux (up to $\sim 5\times$ 
in the optical; $\sim 6\times$ in the X-ray, reaching a peak flux 
$\sim 6 \times 10^{32}d_{kpc}^2 {\rm erg\, s^{-1}}$). These are concentrated
toward the heated side of the companion, but visible at all phases.
Shocks in the wind stripping the companion could be the origin. However, since 
we do not know of such flaring
activity on other well-studied black widow binaries, it is tempting to associate
the activity with the unusually short $P_b$ of J1311. As has been noted \citep[eg.][]{ra93},
low mass stars in close binaries often show enhanced coronospheric activity since
tidal locking enforces rapid rotation, driving dynamo-generated magnetic fields. 
The other main ingredient for high magnetic activity is
convection. This may suggest a low stellar-type mass for the companion. If sub-stellar,
perhaps deep differential heating may drive the convection. A spectroscopic search for
coronal line emission, or spectra during an outburst, could resolve the origin.

	Additional data can greatly improve estimates of the system properties. Simultaneous
multicolor photometry covering many periods and flaring events can help us understand
the temperatures of the various emitting regions. A complete spectroscopic study of
the secondary can place important constraints on the compact object mass and probe
the state of the secondary photosphere. Spectra can also improve knowledge of the compact
object's period, orbital epoch and semi-major axis, each of which are crucial for
decreasing the parameter space to be searched for a $\gamma$-ray pulses.
Of course, a $\gamma$-ray or radio pulsar ephemeris would produce a 
qualitative change in our knowledge of the system properties. We are pursuing all of 
these follow-up investigations.

\begin{figure}[h!!]
\vskip 9.0truecm
\includegraphics{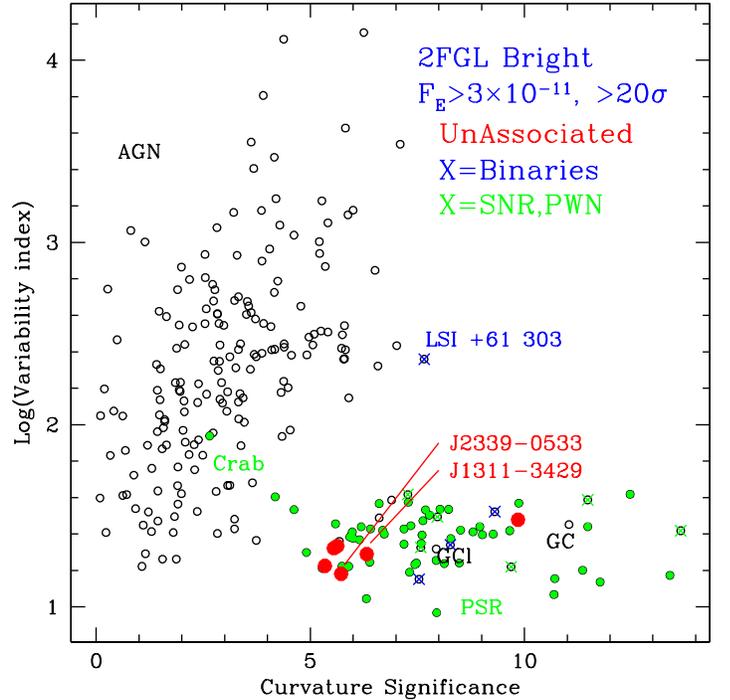}
\begin{center}
\caption{\label{BSpop} 
2FGL bright sources in the spectral curvature - variability plane. Note the excellent 
pulsar(PSR)/blazar(AGN) separation.  Binaries, a globular cluster (GCl), the Galactic Center
(GC) and the unidentified sources belong to the pulsar group.
}
\end{center}
\vskip -0.9truecm
\end{figure}

	We close with a comment on $\gamma$-ray pulsar population in the bright sample. 
Figure 4 shows these sources in the curvature-variability
plane. The pulsar-blazar separation is excellent; the only pulsar in the blazar
region is the Crab, whose PWN contributes the variable power-law component. 
The pulsar region contains a few other source types: high mass binaries, a globular cluster,
SNR, but each may harbor spin-powered pulsars. The
unidentified sources, too, have significant curvature and low variability.  Thus
it is unsurprising that two of the six unidentified sources are good candidates for binary MSP,
although they lack radio detection. The other unidentified sources may be similar or 
new source types. An interesting question is whether J2339$-$0533 and J1311$-$3429
represent the first of a new source class -- radio quiet MSP. It is not
yet clear that we need to appeal to such novelty: for both the optical gives evidence of
a strong evaporative wind. It is entirely possible that radio emission is directed
at Earth, but the pulse is scattered and dispersed to undetectability by this wind. Thus it remains
possible that the MSP $\gamma$-ray beam lies within the radio emission zone -- i.e. that
all $\gamma$-ray MSP are radio detectable.

	Even if these two sources represent a harbinger of a new source class we 
emphasize that this source class must be small, at least for the bright population,
in contrast to some suggestions in the literature \citep{hum05,twc11}. The statistics in
our sample clearly make this case -- of the 41 young pulsars in this set only 17
($1/3$) are radio selected. In contrast there are 12 MSP in the bright sample, all
radio pulsars. Thus even {\it if} all six unidentified sources prove to be radio-quiet
MSP the radio loud subset will be {\it at least} 2/3 of the $\gamma$-ray MSP,
twice the fraction of the young pulsars. Thus, while J1311$-$3429 may have joined the
`black widow' club of pulsars evaporating their companions, it is not yet clear if
there is a `radio quiet MSP' club for it to join. And if such a society does exist, it
is very select, indeed.

\medskip
\medskip
\medskip

This work was supported in part by NASA grant NNX11AO44G.

The SOAR Telescope is a joint project of: Conselho Nacional de Pesquisas Cientificas 
e Tecnologicas CNPq-Brazil, The University of North Carolina at Chapel Hill, 
Michigan State University, and the National Optical Astronomy Observatory.

\end{document}